\documentclass[11pt]{article}
\usepackage{amsmath,amssymb,color,graphics,epsfig,cite}

\usepackage{amsfonts}
\def\td{{\rm total~derivative}}
\newcommand{\hoch}[1]{$\, ^{#1}$}

\makeatletter
\@addtoreset{equation}{section}
\makeatother

\newcommand{\be}{\begin{equation}}
\newcommand{\ee}{\end{equation}}
\newcommand{\bea}{\setlength\arraycolsep{2pt} \begin{eqnarray}}
\newcommand{\eea}{\end{eqnarray}}
\newcommand{\nn}{\nonumber}

\def\cramp{\medmuskip = 2mu plus 1mu minus 2mu}

\def\uncramp{\medmuskip = 4mu plus 2mu minus 4mu}

\def\ft#1#2{{\textstyle{\frac{\scriptstyle #1}{\scriptstyle #2} } }}
\def\fft#1#2{{\frac{#1}{#2}}}
\def\CP{{{\mathbb C}{\mathbb P}}}
\def\0{{\sst{(0)}}}
\def\1{{\sst{(1)}}}
\def\2{{\sst{(2)}}}
\def\3{{\sst{(3)}}}
\def\4{{\sst{(4)}}}
\def\5{{\sst{(5)}}}
\def\6{{\sst{(6)}}}
\def\7{{\sst{(7)}}}
\def\8{{\sst{(8)}}}
\def\sst#1{{\scriptscriptstyle #1}}
\def\oneone{\rlap 1\mkern4mu{\rm l}}

\def\del{{\partial}}

\def\vp{{\varphi}}

\def\td{\tilde}
\def\wtd{\widetilde}
\def\tY{{{\tilde Y}}}

\def\cG{{{\cal G}}}

\def\cH{{{\cal H}}}
\def\im{{{\rm i\,}}}
\def\R{{\mathbb R}}

\def\go{\overset{_{\phantom{.}\circ}}{g}{}}
\def\Do{\overset{_{\phantom{.}\circ}}{D}{}}
\def\po{\overset{_{\phantom{.}\circ}}{\partial}{}}

\begin{document}

\begin{flushright}
\hfill{MI-TH-1613}

\hfill{DAMTP-2016-28}

\end{flushright}

\vspace{25pt}
\begin{center}
{\Large {\bf The Embedding of Gauged STU Supergravity in Eleven Dimensions}}

\vspace{15pt}
{\bf Arash Azizi\hoch{1}, Hadi Godazgar\hoch{2}, Mahdi Godazgar\hoch{2} and 
C.N. Pope\hoch{1,2}}

\vspace{10pt}

\hoch{1} {\it George P. \& Cynthia Woods Mitchell  Institute
for Fundamental Physics and Astronomy,\\
Texas A\&M University, College Station, TX 77843, USA}

\vspace{10pt}

\hoch{2}{\it DAMTP, Centre for Mathematical Sciences,
 Cambridge University,\\  Wilberforce Road, Cambridge CB3 0WA, UK}

 \vspace{20pt}

\vspace{20pt}

\underline{ABSTRACT}
\end{center}

  The consistency of the embedding of four-dimensional $SO(8)$ gauged
${\cal N}=8$ supergravity into eleven-dimensional supergravity, where
the internal directions are compactified on a seven-sphere, was established
by de Wit and Nicolai in the 1980s.  The reduction ansatz for the 
eleven-dimensional metric, and for some of the components of the 
4-form field strength, were found at that time, and recently the
complete expression for the 4-form reduction has been obtained.  The
expressions are quite complicated, and in many practical applications
it would be sufficient to know the ansatz for a subset of the 
four-dimensional fields.  In this paper, we obtain explicit expressions for 
the embedding of the truncation of the full ${\cal N}=8$ gauged theory
to the ${\cal N}=2$ gauged STU supergravity.  This corresponds, in the
bosonic sector, to a 
consistent truncation of the ${\cal N}=8$ supergravity fields to
those that are singlets under the $U(1)^4$ Cartan subalgebra of
$SO(8)$.  This truncation to STU supergravity, which comprises 
${\cal N}=2$ supergravity  
coupled to three vector multiplets, suffices, for example, for
lifting the general 8-charge asymptotically-AdS rotating black holes
to eleven dimensions.  We also give two distinct further truncations
to ${\cal N}=2$ supergravities coupled to single vector multiplets.

\vfill {\footnotesize 
sazizi@physics.tamu.edu,\
h.m.godazgar@damtp.cam.ac.uk,\

m.m.godazgar@damtp.cam.ac.uk,\
pope@physics.tamu.edu.}

\thispagestyle{empty}

\pagebreak

\tableofcontents
\addtocontents{toc}{\protect\setcounter{tocdepth}{2}}



\section{Introduction}

   The idea that the four-dimensional ${\cal N}=8$ gauged $SO(8)$ 
supergravity of de Wit and Nicolai \cite{dewitnicsg} could be obtained
by means of a dimensional reduction of eleven-dimensional supergravity
on $S^7$ dates back to the early 1980s \cite{dufpop,biendeni}.  Originally,
this reduction was discussed just at the level of the linearisation of
the field equations around the Freund-Rubin \cite{frerub} AdS$_4\times S^7$
ground state. 
A reduction in which all the massive Kaluza-Klein towers as well as the
massless ${\cal N}=8$ supergravity subsector is retained can obviously
be consistently extended, in principle, to the full non-linear order.  
However, one cannot in general expect to be able to
perform a consistent truncation of this full theory in which the towers of
massive multiplets are set to zero, leaving only the massless four-dimensional
gauged ${\cal N}=8$ supergravity \cite{gibbonshat}.  The key issue here is
that one would generically expect that in the full theory, non-linear
``currents'' built from powers of the massless fields would act as sources
in the equations of motion of the massive fields that one wishes to set to
zero.  Indeed, this is exactly what would happen in a sphere or coset-space
reduction of any generic theory, and thus a reduction in which just 
a ``massless sector'' (including the gauge bosons of the full isometry group of
the compactifying manifold) was retained would be inconsistent beyond the
linearised level.  The first indication that the $S^7$ reduction of
eleven-dimensional supergravity might be an exception to the general rule
was found in ref.~\cite{dunipowa}, where it is shown that a crucial
trilinear coupling of two $SO(8)$ gauge bosons to a massive spin-2 multiplet
of fields is absent.  This satisfies a first, necessary, condition
for the retained gauge bosons not in fact to act as sources for
the massive fields that one wishes to set to zero.  This happens
because the gauge bosons enter
not only in the metric uplift ansatz but also in the uplift ansatz
for the 4-form field strength of eleven-dimensional supergravity.  This,
together with a certain identity obeyed by the Killing vectors on the
7-sphere, conspires to remove the trilinear couplings that would
otherwise be the first signal of the inconsistency of the 
truncated reduction.

   The possibility of a dimensional reduction of a higher-dimensional
theory on a sphere, in which a finite number of lower-dimensional
fields including the gauge bosons of the isometry group were retained,
was in fact conceived by Pauli in 1953 (in the context of an $S^2$ reduction
of six-dimensional Einstein gravity) 
\cite{pauli,Straumann:2000zc,ORaifeartaigh:1998pk}, but he recognised that
the consistency problems mentioned above would be an obstacle to
realising his idea.  In fact, the $S^7$ reduction of eleven-dimensional
supergravity was the first non-trivial example in which the
idea of a ``Pauli reduction'' \cite{cvgilupo} actually works.

   An indirect, but nevertheless complete, demonstration of the 
consistency of the
$S^7$ reduction was provided by de Wit and Nicolai in 
ref.~\cite{dewitnics7} (see also \cite{nicpil}).
They obtained an explicit expression for the metric uplift ansatz, 
and also partial results for the uplift ansatz for the
4-form field strength.  Their construction, making extensive use of the
supersymmetry of the theory, essentially proves that the reduction
is necessarily a consistent one.  Much more recently, further
work has provided more complete expressions for the 4-form uplift ansatz,
and has also provided further insights into the structure of the 
reduction \cite{dWN13,gengeo,godgodnic,4formans}.

   Although the consistency of the $S^7$ Pauli reduction of eleven-dimensional
supergravity has now been fully established, and the uplift ans\"atze 
for the metric and 4-form field strength are explicitly known, these
expressions are in practice somewhat unwieldy and complicated to use in 
full generality.  
In many cases, when for example lifting a solution of the four-dimensional
gauged supergravity to $D=11$, it may suffice to have explicit expressions
for the uplift ansatz for only a subset of the fields in the
full four-dimensional supergravity theory.  The uplift ansatz for such a 
truncation of the full gauged supergravity theory may be much simpler
and more manageable.  Examples of this kind that have been obtained 
previously include the embedding of four-dimensional 
$SO(4)$-gauged ${\cal N}=4$ supergravity, for which the complete and explicit
bosonic $S^7$ uplift ansatz is given in ref.~\cite{cvluposo4}.  The 
four-dimensional ${\cal N}=4$ gauged theory is of course itself a
consistent truncation of the ${\cal N}=8$ gauged theory.  

   Another
example that has proved to be of considerable utility is given in 
ref.~\cite{tenauthor}.  This describes the embedding of a truncation of the
full $SO(8)$ gauged theory in which only the gauge bosons of the
$U(1)^4$ abelian subgroup of $SO(8)$ are retained.  The truncation
in ref.~\cite{tenauthor} retains also the four-dimensional metric, and three
dilatonic scalar fields.  The associated four-dimensional
theory in this case is not a fully consistent truncation of ${\cal N}=8$
gauged $SO(8)$ supergravity: to be consistent, one should include a
total of six scalar fields rather than just the three that are retained
in ref.~\cite{tenauthor}, comprising three axions as well as the three 
dilatonic
scalars. However, if one restricts attention to four-dimensional
solutions in which the $U(1)$ gauge fields are essentially purely
electric or purely magnetic, but not both, then the three axions can be 
consistently set to zero and the ansatz in ref.~\cite{tenauthor} can then 
be used
in order to lift such solutions to eleven dimensions.\footnote{To be more
precise, the three axions can be consistently set to zero if the
wedge products $F^\alpha\wedge F^\beta$ of the four $U(1)$ field strengths
all vanish.}  The uplift ansatz found in ref.~\cite{tenauthor}, with its
restricted notion of consistency, is sufficient for the purposes of
lifting the static four-charge black hole solutions to eleven dimensions.

  The purpose of the present paper is to obtain explicit expressions
for the bosonic
uplift ansatz for 
the embedding of the fully consistent truncation of the $SO(8)$ gauged 
supergravity to its abelian $U(1)^4$ subsector.  This theory, comprising 
the metric, four $U(1)$ gauge fields, and the three dilatons and three
axions mentioned above, is the bosonic sector of the ${\cal N}=2$
supersymmetric gauged STU supergravity theory.  Its field content
comprises the ${\cal N}=2$ supergravity multiplet coupled to three
vector multiplets.  The inclusion of the additional three axionic
scalar fields makes the uplift ansatz considerably more complicated than
the restricted uplift ansatz that was found in ref.~\cite{tenauthor}.  
In fact an attempt to construct the ansatz for the embedding of
the bosonic sector of the gauged STU supergravity was made in ref.~\cite{cvlupostutry}, but at that time only the partial results for the
4-form uplift ansatz that had been obtained in ref.~\cite{dewitnics7}
were available, and the results in ref.~\cite{cvlupostutry} are for that
reason incomplete.  With the recent advances in the construction of
the internal 4-form uplift ansatz for the $SO(8)$ gauged supergravity
that have been achieved in refs.~\cite{dWN13,testing,godgodnic},
we are now in a position to complete the job that was left unfinished in 
ref.~\cite{cvlupostutry}.  

The ansatz that we shall present in this paper is considerably more complicated than the restricted one without the
three axionic scalars that was obtained in ref.~\cite{tenauthor}.  It is, 
however, still considerably simpler, for practical purposes, than the
complete ansatz for the reduction to the full ${\cal N}=8$ gauged $SO(8)$ 
supergravity theory.  The ansatz we obtain allows one to lift,
for example, any black hole solution of gauged ${\cal N}=8$ supergravity carrying abelian charges to eleven dimensions.  The most general such solution would have four electric and four magnetic charges (although global symmetries could be used in order to rotate to a duality complexion where a total of 5 independent non-zero charges remain).

\section{Gauged STU supergravity}

The gauged STU supergravity is a consistent truncation of the ${\cal N} = 8$ $SO(8)$ gauged theory with ${\cal N} = 2$ supersymmetry and residual gauge group $U(1)^4$, the maximal abelian subgroup of $SO(8)$ \cite{duffliu, tenauthor}.  In particular, the truncation leaves a total of six non-trivial scalars, three dilatons and three axions parameterised by $\lambda_i$ and $\sigma_i$, respectively, with $i=1,2,3,$ and four $U(1)$ gauge fields $A^{\alpha}$ ($\alpha=1,2,3,4$) with associated field strengths $F^{\alpha}$.  The bosonic Lagrangian, in the notation of ref.~\cite{cvlupostutry}, is
\begin{equation}
 { L} = R\, {*\oneone} + {L}_{KinS} - V\, {*\oneone} + {L}_{KinA} + {L}_{CS},
\end{equation}
where the scalar kinetic terms are simply \cite{cvlupostutry}
\begin{equation}
 { L}_{KinS} = -\frac{1}{2} \sum_{i} \Bigg( (\partial \lambda_i)^2 + 
\sinh^2{\lambda_i} \, (\partial \sigma_i)^2 \Bigg)\,{*\oneone}
\end{equation}
and the scalar potential 
\begin{equation}
 V = - 4\, g^2 \sum_{i} (Y_i^2 + \widetilde{Y}_{i}^2),
\end{equation}
where \cite{cvlupostutry}
\begin{equation}
\cosh{\lambda_i} = \frac{1}{2}(Y_i^2 + \widetilde{Y}_i^2), 
\qquad \cos{\sigma_i}\, \sinh{\lambda_i} = \frac{1}{2}(Y_i^2 - 
\widetilde{Y}_i^2). 
\end{equation}
In addition, we define \cite{cvlupostutry}
\begin{equation}
 \sin{\sigma_i}\, \sinh{\lambda_i} = b_i.
\end{equation}

  It will sometimes be more convenient to use an $SL(2,\R)$ parameterisation
for the scalar fields rather than an $SO(2,1)$ parameteristion,
by defining dilaton/axion pairs $(\varphi_i,\chi_i)$ by
\be
e^{\varphi_i}= \cosh\lambda_i + \sinh\lambda_i\, \cos\sigma_i\,,\qquad
\chi_i\, e^{\varphi_i}= \sinh\lambda_i\, \sin\sigma_i\,,\label{dilax}
\ee
in terms of which the scalar kinetic Lagrangian becomes
\be
 { L}_{KinS} = -\frac{1}{2} \sum_{i} \Bigg( (\partial \varphi_i)^2 + 
e^{2\varphi_i} \, (\partial \chi_i)^2 \Bigg)\,{*\oneone}\,.
\ee
The quantities $Y_i$, $\wtd Y_i$ and $b_i$ defined above are now given by
\be
Y^2_i= e^{\varphi_i}\,,\qquad 
\wtd Y_i^2 = e^{-\varphi_i} + \chi_i^2\, e^{\varphi_i}\,,\qquad
b_i=\chi_i\, e^{\varphi_i}\,.
\ee

The kinetic terms for the gauge fields ${\cal L}_{KinA}$ and the Chern-Simons 
terms ${\cal L}_{CS}$, given by equations (36) and (38) of 
ref.~\cite{cvlupostutry}, are
\bea
{\cal L}_{KinA} &=& 
-\ft12|W|^{-2}\, \Big[P_0\,\Big(\tY_1^2\, \tY_2^2\, \tY_3^2\,
{*F_\2^1}\wedge F_\2^1 +
\tY_1^2\, Y_2^2\, Y_3^2\, {*F_\2^2}\wedge F_\2^2 \nn\\
&&\qquad\qquad\quad + Y_1^2\, \tY_2^2\, Y_3^2\, {*F_\2^3}\wedge F_\2^3 
+ Y_1^2\, Y_2^2\, \tY_3^2\, {*F_\2^4}\wedge F_\2^4\Big)\nn\\
&& \qquad\qquad+ 2 P_1\, b_2\, b_3\, (\tY_1^2\, {*F_\2^1}\wedge F_\2^2 
-Y_1^2\, {*F_\2^3}\wedge F_\2^4)\nn\\
&& \qquad\qquad + 2 P_2\, b_1\, b_3\, (\tY_2^2\, {*F_\2^1}\wedge F_\2^3 
-Y_2^2\, {*F_\2^2}\wedge F_\2^4)\nn\\
&& \qquad\qquad + 2 P_3\, b_1\, b_2\, (\tY_3^2\, {*F_\2^1}\wedge F_\2^4 
-Y_3^2\, {*F_\2^2}\wedge F_\2^3)\Big]\,\label{lkin}
\eea
and
\bea
{\cal L}_{CS} &=& -|W|^{-2}\, \Big[ b_1\, b_2\, b_3\, 
\Big(\tY_1^2\, \tY_2^2\, \tY_3^2\, F_\2^1\wedge F_\2^1 +
\tY_1^2\, Y_2^2\, Y_3^2\, F_\2^2\wedge F_\2^2\nn\\
&& \qquad\qquad\qquad\quad + Y_1^2\, \tY_2^2\, Y_3^2\, F_\2^3\wedge F_\2^3 
+ Y_1^2\, Y_2^2\, \tY_3^2\, F_\2^4\wedge F_\2^4\Big)\nn\\
&&
\qquad\qquad
+b_1\, ( P_0 + 2 b_2^2\, b_3^2)\, (\tY_1^2\, F_\2^1\wedge F_\2^2 
-Y_1^2\, F_\2^3\wedge F_\2^4)\nn\\
&&\qquad\qquad
+ b_2\, (P_0 + 2 b_1^2\, b_3^2)\, (\tY_2^2\, F_\2^1\wedge F_\2^3 
-Y_2^2\, F_\2^2\wedge F_\2^4)\nn\\
&&
\qquad\qquad
+b_3\, (P_0 + 2 b_1^2\, b_2^2)\, (\tY_3^2\, F_\2^1\wedge F_\2^4 
-Y_3^2\, F_\2^2\wedge F_\2^3)\Big]\,,\label{lcs}
\eea
 where
\bea
&&P_0 \equiv 1 +b_1^2 + b_2^2 +b_3^2\,,\quad
W \equiv P_0 - 2\im\, b_1\, b_2\, b_3\,,\nn\\
&&
P_1\equiv  1 -b_1^2 + b_2^2 +b_3^2\,,\quad
P_2\equiv  1 +b_1^2 - b_2^2 +b_3^2\,,\quad
P_3\equiv  1 +b_1^2 + b_2^2 -b_3^2\,.\label{WPdef}
\eea

\section{Complete ans\"atze for ${\cal N}=8$ gauged $SO(8)$ supergravity} \label{sec:ansaetze}

In this section we summarise briefly the uplift ans\"atze for the full 
${\cal N}=8$ gauged $SO(8)$ supergravity \cite{dWNW85, dWN13, testing, godgodnic, 4formans}.~\footnote{We simply state the uplift ans\"atze here and do not 
give an account of how they are derived. For such an account, the reader is 
encouraged to consult Refs.~\cite{gengeo, godgodnic, 4formans}.} We remark 
that the ans\"atze presented below are not unique, and in fact 
there are many possible ways of writing them 
down \cite{Varela:2015ywx,Kruger:2016agp}.  Of course, these ans\"atze 
should all be equivalent and the consistency of both the $D=11$ and 
${\cal N}=8$ theories as well as the reduction will guarantee that this 
will indeed be the case.  Here, we present the ans\"atze in what
we believe to be their simplest known form.

The metric ansatz \cite{dWNW85}
\begin{equation}
   \Delta^{-1} g^{mn}(x,y) = \frac{1}{8} K^{m\,IJ}(y) K^{n\,KL}(y) 
   \Big[\left( u^{MN}{}_{IJ} + v^{MNIJ} \right) 
    \left( u_{MN}{}^{KL} + v_{MNKL} \right) \Big](x) \label{ans:metric} 
\end{equation}
has been known for some time and is in fact applied in ref.~\cite{cvlupostutry} to determine the internal metric for the full gauged STU supergravity.  The recent progress concerns uplift ans\"atze for components of the field strength $F_{MNPQ}$ (or equivalently the three form potential $A_{MNP}$), \textit{viz.}
\begin{equation} \label{Fcomp}
 F_{mnpq}, \quad F_{\mu mnp}, \quad F_{\mu \nu mn}, \quad F_{\mu \nu \rho m}\quad  \text{and}\quad  F_{\mu \nu \rho \sigma}.
\end{equation}
Note that the components as presented above do not quite correspond in a direct manner to the notation used in ref.~\cite{cvlupostutry}.  This is because, in line with the whole spirit of a Kaluza-Klein reduction, all fields in ref.~\cite{cvlupostutry} are defined with the Kaluza-Klein gauge fields included; i.e.\ 
the seven-dimensional coordinate differentials $dy^m$ always appear in the 
Kaluza-Klein covariantised form
\begin{equation}
dy^m \longrightarrow 
\Big(dy^m - K^{m IJ}(y)\,  A_{\mu}^{IJ}(x)\,  dx^\mu \Big)\,.
\end{equation}
The Kaluza-Klein redefinition above is easy to account for and we 
shall do so when presenting the full set of uplift ans\"atze 
(see section \ref{sec:uplift}).

The $F_{mnpq}$ components of the field strength can be obtained simply from the ansatz for $A_{mnp}$\cite{dWN13, testing, godgodnic}
\begin{equation}
A_{mnp}(x,y) 
= - \frac{i\sqrt{2}}{96} \, \Delta g_{pq} \, K_{mn}^{IJ}(y) K^{q\, KL}(y) 
  \Big[ \left( u^{MN}{}_{IJ} -  v^{MNIJ} \right) \left( u_{MN}{}^{KL} + v_{MNKL} \right) \Big](x).
 \label{ans:flux}
\end{equation}
In particular,
\begin{equation}
 F_{mnpq} = 4 \partial_{[m}A_{npq]}.
\end{equation}
Similarly,
\begin{equation}
 F_{\mu mnp} = \partial_{\mu} A_{mnp} - 3 \partial_{[m} A_{|\mu|np]}
\end{equation}
with $A_{mnp}$ as determined by equation \eqref{ans:flux} 
and $A_{\mu mn}$ given by the following exact vector 
ansatz \cite{godgodnic}
\begin{equation}
 A_{\mu mn}(x,y) = \frac{1}{24} K_{mn}{}^{IJ}(y) A_{\mu\, IJ}(x) - \frac{\sqrt{2}}{4} A_{mnp}(x,y) K^{p\, IJ}(y) A_{\mu}{}^{IJ}(x).
\end{equation}

The ansatz for $F_{\alpha \beta a b}$ \cite{4formans} is particularly simple
\begin{equation} \label{eq:F22s7} 
 e_{\mu}{}^{\alpha} e_{\nu}{}^{\beta} e_{m}{}^{a} e_{n}{}^{b} F_{\alpha \beta a b} = \frac{\sqrt{2}}{8} \Bigg(K_{mn}{}^{IJ}(y)\, \cG_{\mu \nu\, IJ}(x) - 12 \Delta^{-1} A_{mnp}(x,y)\, K^{p}{}_{IJ}(y)\, \cH_{\mu \nu}{}^{IJ}(x) \Bigg),
\end{equation}
where $\cG_{\mu \nu\, IJ}$ and $\cH_{\mu \nu}{}^{IJ}$ are covariantised field strengths of the electric and magnetic vectors and $e_{\mu}{}^{\alpha}$ and $e_{m}{}^{a}$ are defined by
\begin{equation}
 E_{M}{}^{A} = \begin{pmatrix}
         e_{\mu}{}^{\alpha} & B_{\mu}{}^{p} e_{p}{}^{a} \\
         0 & e_{m}{}^{a}
               \end{pmatrix}.
\end{equation}
The Kaluza-Klein vectors $B_{\mu}{}^{m}$ in the STU truncation are given 
by the abelian $U(1)^4$ truncation of $K^{mIJ} \,A_{\mu}^{IJ}$. 
In particular, from the above ansatz we deduce that $F_{\alpha \beta a b}$ vanishes for four-dimensional solutions with vanishing vector expectations values.

The ansatz for $F_{\alpha \beta \gamma a}$ is perhaps the most involved \cite{4formans}
\begin{align} \label{eq:F31ansatzs7}
 &e_{\mu}{}^{\alpha} e_{\nu}{}^{\beta} e_{\rho}{}^{\gamma} e_{m}{}^{a} F_{ \alpha \beta \gamma a} \notag \\
 & \hspace{5mm} = - \frac{\sqrt{2}}{48} \eta_{\mu \nu
    \rho}{}^{\sigma} K^{n \; IJ}(y)\, K_{mn}{}^{KL}(y)\, \left[\left(u^{ij}{}_{IJ} +
    v^{ij IJ} \right) \left( u^{kl}{}_{KL}- v^{kl KL} \right)\right](x)  \notag
  \\[3mm]
  & \hspace{9mm}\times\left( \hat{{\cal V}}^{{\cal M}}{}_{ij} \partial_\sigma
    \hat{{\cal V}}_{{\cal M} \, kl} - 2 \sqrt{2} m_7 A_{\sigma}{}^{MN}
    \hat{{\cal V}}^{MP}{}_{ij} \hat{{\cal V}}_{NP\, kl} - 2 \sqrt{2} m_7
    A_{\sigma}{}^{MN} \hat{{\cal V}}^{MP}{}_{kl} \hat{{\cal V}}_{NP\,
      ij}\right)(x),
\end{align}
where $\hat{{\cal V}}_{{\cal M}\, ij}$ is usual four-dimensional $E_{7(7)}/SU(8)$ coset element parameterising the 70 scalars.

The ansatz for the $F_{\mu \nu mn}$ components of the four-form are
\begin{align}
 F_{\mu \nu mn} &= E_{\mu}{}^{A} E_{\nu}{}^{B} E_{m}{}^{C} E_{n}{}^{D} F_{A B C D} \notag \\
                &= e_{m}{}^{a} e_{n}{}^{b} e_{\mu}{}^{\alpha} e_{\nu}{}^{\beta}  F_{\alpha \beta a b} + 
                2 B_{\nu}{}^{p} F_{\mu mnp} 
                - B_{\mu}{}^{p} B_{\nu}{}^{q} F_{mnpq}, 
\end{align}
where the first term is given in equation \eqref{eq:F22s7}. Similarly, the $ F_{\mu \nu \rho m}$ are given by equation \eqref{eq:F31ansatzs7} and appropriate contractions of the $F_{\mu \nu mn}$, $F_{\mu mnp}$ and $F_{mnpq}$ components with the Kaluza-Klein vectors $B_{\mu}{}^{m}.$~\footnote{Note that in ref.~\cite{4formans}, the components on the left-hand side of equations \eqref{eq:F22s7} and \eqref{eq:F31ansatzs7} are denoted by $F_{\mu \nu mn}$ and $F_{\mu \nu \rho m}$, respectively.} 

On the other hand, the \emph{conjectured} ansatz for the Freund-Rubin term~\footnote{There are other established ans\"atze for the Freund-Rubin term (see ref.~\cite{4formans}). However, it has not yet been possible to show that those ans\"atze have the following simple form.} is perhaps the simplest and most elegant 
\cite{4formans}~\footnote{In this paper, we use the conventions of ref.~\cite{cvlupostutry}, which are related to the conventions of ref.~\cite{4formans} as follows: $g^{\textup{ref.~\cite{4formans}}} = \sqrt{2} \, g^{\textup{ref.~\cite{cvlupostutry}}}$ and $V^{\textup{ref.~\cite{4formans}}} = 1/2 \, V^{\textup{ref.~\cite{cvlupostutry}}}$.} 
\begin{equation} \label{frconjans}
 \mathfrak{f}_{FR}= -\frac{i}{24} \eta^{\alpha \beta \gamma \delta} F_{\alpha \beta \gamma \delta} = -\frac{m_7}{96\sqrt{2}g^2}\Bigg( V(x) - \frac{g^2}{6}\Big(Q^{ijkl}(x)\hat{\Sigma}_{ijkl}(x,y) + \text{h.c.}\Big) \Bigg),
\end{equation}
where $V(x)$ is the four-dimensional scalar potential, $Q^{ijkl}$ is simply given in terms of the $T$-tensor
\begin{equation}
  \label{eq:Qijkl}
  Q^{ijkl} = \frac{3}{4} A_{2\;m}{}^{n[ij} A_{2\;n}{}^{kl]m} -
  A_1{}^{m[i} A_{2\;m}{}^{jkl]}
\end{equation}
and complex self-dual tensor
\begin{equation}\label{Sigmahat}
  \hat \Sigma_{ijkl} (x,y) =  \left(   u_{ij}{}^{IJ} u_{kl}{}^{KL} - v_{ij\;IJ} v_{kl\;KL} \right)(x)\ K^{m\, [IJ}(y) K_{m}{}^{KL]}(y).
\end{equation}

Note that the only components of the four-form field strength for which 
an ansatz is not given in ref.~\cite{cvlupostutry} are the ones related 
to the \emph{internal} components $A_{mnp}$ of $A_{MNP}$, 
which, as explained above, are given by 
the ansatz in equation \eqref{ans:flux}, after including 
the Kaluza-Klein redefinitions explained above and in 
section \ref{sec:uplift}.

\section{Derivation of $A_{mnp}$} \label{sec:A}

In this section, we outline the derivation of the purely internal (7-sphere)
components $A_{mnp}$ using ansatz \eqref{ans:flux}, which we repeat here for 
convenience
\begin{equation} \label{Amnp}
 A_{mnp} 
= - \frac{i\sqrt{2}}{96} \, \Delta g_{pq}\, K_{mn}^{IJ} K^{q\, KL}
 \left( u^{MN}{}_{IJ} -  v^{MNIJ} \right) \left( u_{MN}{}^{KL} + v_{MNKL} \right).
\end{equation}
As is clear from the ansatz above, the first step is to compute the $u$ 
and $v$ tensors. These tensors have already been computed in 
ref.~\cite{cvlupostutry} (see also ref.~\cite{Lu:2014fpa}), but here, 
we express them in terms of $U(1)^4$-invariant tensors. In this way the 
$u$ and $v$ are expressed more covariantly with respect to the $U(1)^4$ 
symmetry. 

Without loss of generality, we can choose 
$U(1)^4 \subset SO(8)$ to act on the index pairs $\{12\}, \{34\}, \{56\}$ 
and $\{78\},$ where, for example the first $U(1)$ rotates $1$ and $2$ into each other.
Introducing
\be
(X^{IJKL})_{I'J'K'L'} = 4!\, \delta^{IJKL}_{I'J'K'L'}\,,
\ee
we define $C^i_\pm$ as
\be
C^1_\pm = X^{1234}\pm X^{5678}\,,\qquad
C^2_\pm = X^{1256} \pm X^{3478}\,,\qquad
C^3_\pm = X^{1278} \pm X^{3456}\,.
\ee
It is clear that $C^i_\pm$ are the unique rank-4 $U(1)^4$-invariant tensors, up to redefinitions of the embedding of  $U(1)^4 \subset SO(8)$. Furthermore, there are also four rank-2 invariant tensors
\begin{equation}
 F^{(1)}_{IJ} = 2 \, \delta^{12}_{IJ}, \qquad   F^{(2)}_{IJ} = 2\, \delta^{34}_{IJ}, \qquad   F^{(3)}_{IJ} = 2\, \delta^{56}_{IJ}, \qquad  F^{(4)}_{IJ} = 2 \, \delta^{78}_{IJ}.   
\end{equation}

With the 28 Killing vectors of the $SO(8)$ symmetry of the round $S^7$ written
in the spinor representation, i.e.\ with 
\be
K^{IJ}= \ft12 (\Gamma_{ab})^{IJ}\, \Big(x^a\fft{\del}{\del x^b} - 
                    x^b\, \fft{\del}{x^a}\Big)\,,
\ee
we then define the following vectors and tensors:
\begin{gather}
\xi_m^{(i)} = \ft1{16} C^{(i)}_{+IJKL}\, K_{mn}^{IJ}\, K^{n KL}\,, \qquad
\xi_{mn}^{(i)} = -\ft1{16} C^{(i)}_{+IJKL}\, K_m^{IJ}\, K_n^{KL}\,,\nn\\[2mm]
S^{(i)}_{mnp} = \ft1{16} C^{(i)}_{-IJKL}\, K_{mn}^{IJ}\, K_p^{KL}\,,\nn\\[2mm]
F^{(\alpha)}_m = F^{(\alpha)}_{IJ}\, K^{IJ}_m\,, \qquad
F^{(\alpha)}_{mn} = F^{(\alpha)}_{IJ}\, K_{mn}^{IJ}\,,
\label{so7tensors}
\end{gather}
where
\begin{equation}
 K_{mn}{}^{IJ} = K_{[mn]}{}^{IJ} = -\frac{1}{m_7} \Do_{m} K_{n}{}^{IJ}.
\end{equation}
Indices on $K_{m}{}^{IJ}$ and $K_{mn}{}^{IJ}$ as well as all $U(1)^4$-invariant tensors are raised and lowered with the background (inverse) metric $\go^{mn}$ and $\go_{mn}$, respectively.

The 56-bein $\mathcal{V}= \mathcal{V}_1 \mathcal{V}_2 \mathcal{V}_3$, where 
\begin{equation}
 \mathcal{V}_i = \begin{pmatrix}
                  u^{(i)}{}_{ij}{}^{IJ} & v^{(i)}{}_{ij KL} \\
                  v^{(i)}{}^{kl IJ} & u^{(i)}{}^{kl}{}_{KL}
                 \end{pmatrix}.
\end{equation}
In symmetric gauge each 56-bein is expressed as 
 \begin{equation}
 \mathcal{V}_i = \exp \begin{pmatrix}
                  0  & \phi^{(i)} \\
                  \bar{\phi}^{(i)} & 0
                 \end{pmatrix},
\end{equation}
where $ \bar{\phi}^{(i)} $ is the complex conjugate of  $\phi^{(i)}.$ 
In terms of the $U(1)^4$-invariant tensors above, the scalar expectation values are 
\begin{equation}
 \phi^{(i)} = -2\lambda_{i} \left( \cos \sigma_{i} \, C^{i}_+ - i \, \sin{\sigma_{i}} \, C^i_{-} \right).
\end{equation}
Hence by exponentiation,
\begin{align}
 u^{(i)}{}_{IJ}{}^{KL} &= \delta^{KL}_{IJ} + \frac{1}{4} \left( \cosh \frac{\lambda_i}{2} -1 \right) \left(C^{i}_{+} C^{i}_{+ } \right)_{IJKL}, \\
 v^{(i)}{}_{IJKL} &=  -\frac{1}{2} \sinh{ \frac{\lambda_i}{2}}  \left( \cos \sigma_{i} \, C^{i}_+ - i \, \sin{\sigma_{i}} \, C^i_{-} \right)_{IJKL}.
\end{align}

The $u$ and $v$ matrices corresponding to the full 56-bein $\mathcal{V}$ can then be found from
\begin{align}
 u_{IJ}{}^{KL} &= u^{(1)}{}_{IJ}{}^{MN} \left( u^{(2)}{}_{MN}{}^{PQ} u^{(3)}{}_{PQ}{}^{KL} + v^{(2)}{}_{MNPQ} v^{(3)}{}^{PQKL} \right) \notag \\
 & \quad + v^{(1)}{}_{IJMN} \left(v^{(2)}{}^{MNPQ}  u^{(3)}{}_{PQ}{}^{KL} + u^{(2)}{}^{MN}{}_{PQ} v^{(3)}{}^{PQKL} \right), \\
 v^{IJKL} &= v^{(1)}{}^{IJMN} \left( u^{(2)}{}_{MN}{}^{PQ} u^{(3)}{}_{PQ}{}^{KL} + v^{(2)}{}_{MNPQ} v^{(3)}{}^{PQKL} \right) \notag \\
 & \quad + u^{(1)}{}^{IJ}{}_{MN} \left(v^{(2)}{}^{MNPQ}  u^{(3)}{}_{PQ}{}^{KL} + u^{(2)}{}^{MN}{}_{PQ} v^{(3)}{}^{PQKL} \right).
\end{align}

Now that we have the $u$ and $v$ tensors, we can compute $\Delta^{-1} g^{pq} A_{mnp}$ by contracting them with $K_{mn}^{IJ}$ and $K^{q\, KL}$ as prescribed by the formula \eqref{Amnp}.  In terms of the $U(1)^4$-invariant vectors and tensors defined in equations \eqref{so7tensors},\footnote{Recall that indices on $U(1)^4$-invariant tensors are raised/lowered using the background (inverse) metric on the round seven-sphere, $\go^{mn}$/$\go_{mn}$.}
\begin{align}
 \Delta^{-1} g^{pq} A_{mnp} = \sum_{i=1}^{3}\, b_{i}\, S^{(i)q}{}_{mn} + \frac{1}{64} \sum_{\alpha, \beta=1}^{4} f_{\alpha \beta}\ F^{(\alpha) q}\, F^{(\beta)}_{mn},
\end{align}
where, defining 
\begin{equation}
 y^{2}_{i} = Y_i^2 - \widetilde{Y}_i^2, \qquad \tilde{y}^{2}_{i} = Y_i^2 
+ \widetilde{Y}_i^2,
\end{equation}
\begin{align}
 f_{11} &= y_{1}^2 (b_3 y_2^2 + b_2 y_3^2) + b_1 y_2^2 y_3^2 - 4 b_1 b_2 b_3, \qquad
 f_{22} = - y_1^2 (b_3 y_2^2 + b_2 y_3^2) + b_1 y_2^2 y_3^2 - 4 b_1 b_2 b_3, \nn \\
 f_{33} &= - y_1^2 (b_3 y_2^2 - b_2 y_3^2) - b_1 y_2^2 y_3^2 - 4 b_1 b_2 b_3, \quad \;
 f_{44} = y_1^2 (b_3 y_2^2 - b_2 y_3^2) - b_1 y_2^2 y_3^2 - 4 b_1 b_2 b_3, \nn \\
 f_{12} &=  - \tilde{y}_{1}^2 (b_3 y_2^2 + b_2 y_3^2) + b_1 \tilde{y}_2^2 \tilde{y}_3^2 - 4 b_1, \qquad \quad
 f_{21} = \tilde{y}_{1}^2 (b_3 y_2^2 + b_2 y_3^2) + b_1 \tilde{y}_2^2 \tilde{y}_3^2 - 4 b_1, \nn \\
 f_{13} &= - \tilde{y}_{2}^2 (b_3 y_1^2 + b_1 y_3^2) + b_2 \tilde{y}_1^2 \tilde{y}_3^2 - 4 b_2, \qquad \quad
 f_{31} = \tilde{y}_{2}^2 (b_3 y_1^2 + b_1 y_3^2) + b_2 \tilde{y}_1^2 \tilde{y}_3^2 - 4 b_2, \nn \\
 f_{14} &= - \tilde{y}_{3}^2 (b_2 y_1^2 + b_1 y_2^2) + b_3 \tilde{y}_1^2 \tilde{y}_2^2 - 4 b_3, \qquad \quad
 f_{41} =  \tilde{y}_{3}^2 (b_2 y_1^2 + b_1 y_2^2) + b_3 \tilde{y}_1^2 \tilde{y}_2^2 - 4 b_3, \nn \\
 f_{23} &= \tilde{y}_{3}^2 (b_2 y_1^2 - b_1 y_2^2) - b_3 \tilde{y}_1^2 \tilde{y}_2^2 + 4 b_3, \qquad \quad \ \ \,
 f_{32} = - \tilde{y}_{3}^2 (b_2 y_1^2 - b_1 y_2^2) - b_3 \tilde{y}_1^2 \tilde{y}_2^2 + 4 b_3, \nn \\
 f_{24} &= \tilde{y}_{2}^2 (b_3 y_1^2 - b_1 y_3^2) - b_2 \tilde{y}_1^2 \tilde{y}_3^2 + 4 b_2, \qquad \quad \ \ \,
 f_{42} = -\tilde{y}_{2}^2 (b_3 y_1^2 - b_1 y_3^2) - b_2 \tilde{y}_1^2 \tilde{y}_3^2 + 4 b_2, \nn \\
 f_{34} &= \tilde{y}_{1}^2 (b_3 y_2^2 - b_2 y_3^2) - b_1 \tilde{y}_2^2 \tilde{y}_3^2 + 4 b_1, \qquad \quad \ \ \,
 f_{43} = - \tilde{y}_{1}^2 (b_3 y_2^2 - b_2 y_3^2) - b_1 \tilde{y}_2^2 \tilde{y}_3^2 + 4 b_1.
\end{align}
The $U(1)^4$-invariant vectors and tensors are given explicitly in terms of a set of adapted coordinates on $S^7$ in appendix \ref{app:coords}. Thus, all that is left to do in order to find $A_{mnp}$ is to contract $\Delta g_{pq},$ which we know from the metric ansatz \cite{cvlupostutry}, with the expression found above written in adapted coordinates, i.e.
\begin{equation}
 A_{mnp} =\Delta g_{pq} \left(\sum_{i=1}^{3}\, b_{i}\, S^{(i)q}{}_{mn} + \frac{1}{64} \sum_{\alpha, \beta=1}^{4} f_{\alpha \beta}\ F^{(\alpha) q}\, F^{(\beta)}_{mn}\right).
\end{equation}
The resulting expression is~\footnote{The factor of $g^{-3}$ is introduced in 
the expressions below because we are using dimensionless coordinates 
on the unit sphere.}

\cramp
\bea
A_{\alpha 56} \,d\mu_\alpha&=& \fft{b_1}{2\, \Xi \, g^3}\,
    \Big[\mu_1^2\, W_2 \,d(\mu_2^2) - 
   \mu_2^2 \,W_1 \,d(\mu_1^2) 
   -\mu_1^2 \,\mu_2^2\, (Y_2^2 \widetilde Y_2^2 \,d\alpha_2 +
   Y_3^2 \widetilde Y_3^2 \,d\alpha_3)\Big]\,,\nn\\
A_{\alpha 78} \, d\mu_\alpha&=& \fft{b_1}{2\, \Xi \, g^3}\,
    \Big[\mu_4^2 \,W_3\, d(\mu_3^2) - 
         \mu_3^2\, W_4 \, d(\mu_4^2)
   +\mu_3^2 \, \mu_4^2\,  (Y_2^2 \widetilde Y_2^2 \,d\alpha_2 -
   Y_3^2 \widetilde Y_3^2 \, d\alpha_3)\Big]\,,\nn\\
A_{\alpha 57} \, d\mu_\alpha&=& \fft{b_2}{2\, \Xi \, g^3}\,
\Big[\mu_1^2\, W_3 \,d(\mu_3^2) 
          - \mu_3^2 \, W_1\, d(\mu_1^2)
   -\mu_1^2 \, \mu_3^2\,  (Y_1^2 \widetilde Y_1^2 \, d\alpha_1 +
   Y_3^2 \widetilde Y_3^2 \, d\alpha_3)\Big]\,,\nn\\
A_{\alpha 68} \, d\mu_\alpha&=& \fft{b_2}{2\, \Xi \, g^3}\,
  \Big[\mu_4^2\, W_2 \,d(\mu_2^2) 
          - \mu_2^2\, W_4 \, d(\mu_4^2)
   +\mu_2^2 \, \mu_4^2 \, (Y_1^2 \widetilde Y_1^2 \, d\alpha_1 -
   Y_3^2 \widetilde Y_3^2 \, d\alpha_3)\Big]\,,\nn\\
A_{\alpha 58} \, d\mu_\alpha&=& \fft{b_3}{2\, \Xi \, g^3}\,
   \Big[\mu_1^2 \, W_4 \, d(\mu_4^2) 
         - \mu_4^2\, W_1 \, d(\mu_1^2)
   -\mu_1^2 \, \mu_4^2 \, (Y_1^2 \widetilde Y_1^2 \, d\alpha_1 +
   Y_2^2 \widetilde Y_2^2 \, d\alpha_2)\Big]\,,\nn\\
A_{\alpha 67} \, d\mu_\alpha&=& \fft{b_3}{2\, \Xi \, g^3}\, 
   \Big[\mu_3^2 \, W_2 \, d(\mu_2^2) 
            - \mu_2^2 \, W_3 \, d(\mu_3^2)
   +\mu_2^2 \, \mu_3^2 \, (Y_1^2 \widetilde Y_1^2 \, d\alpha_1 -
   Y_2^2 \widetilde Y_2^2 \, d\alpha_3)\Big]\,,
   \label{A}
\eea
\uncramp
where
\bea
W_1 &=& \widetilde Y_2^2 \widetilde Y_3^2 \, \mu_2^2 +
    \widetilde Y_1^2 \widetilde Y_3^2\, \mu_3^2 + 
  \widetilde Y_1^2 \widetilde Y_2^2 \, \mu_4^2\,,\nn\\
W_2 &=& Y_2^2  Y_3^2 \, \mu_1^2 +
    \widetilde Y_1^2 Y_2^2\, \mu_3^2 + 
  \widetilde Y_1^2 Y_3^2 \, \mu_4^2\,,\nn\\
W_3 &=& Y_1^2 Y_3^2 \, \mu_1^2 +
   Y_1^2 \widetilde Y_2^2\, \mu_2^2 + 
  \widetilde Y_2^2 Y_3^2 \, \mu_4^2\,,\nn\\
W_4 &=& Y_1^2 Y_2^2 \, \mu_1^2 +
    Y_1^2 \widetilde Y_3^2\, \mu_2^2 + 
  Y_2^2 \widetilde Y_3^2 \, \mu_3^2\, \label{Widef}
\eea
and
\be
\alpha_1 = \mu_1^2 + \mu_2^2\,\qquad \alpha_2=\mu_1^2+\mu_3^2\,,
\qquad \alpha_3 = \mu_1^2 + \mu_4^2\,.
\ee
Note that the $W_\alpha$ defined above are related to the 
$Z_\alpha$ defined in ref.~\cite{cvlupostutry} as follows
\be
W_\alpha = Z_\alpha -\mu_\alpha^2 \,.  \label{Zdef}
\ee
The coordinate indices $(5,6,7,8)$ on the $A_{mnp}$ refer to the
directions $(\phi_1, \phi_2,\phi_3,\phi_4)$ respectively, so $A_\3=
  A_{\alpha 56}\, d\mu_\alpha\wedge d\phi_1\wedge d\phi_2+\cdots$.  
  
The function $\Xi$ is defined in equation (21) of ref.~\cite{cvlupostutry} as 
follows
\begin{align}
 \Xi \, =\ &Y_1^2 Y_2^2 Y_3^2 \mu_1^4 + Y_1^2 \widetilde{Y}_2^2 
\widetilde{Y}_3^2 \mu_2^4 + \widetilde{Y}_1^2 Y_2^2 \widetilde{Y}_3^2 \mu_3^4 
 + \widetilde{Y}_1^2 \widetilde{Y}_2^2 Y_3^2 \mu_4^4 \nn \\
 & + (Y_2^2 \widetilde{Y}_2^2 + Y_3^2 \widetilde{Y}_3^2) (Y_1^2 \mu_1^2 
\mu_2^2 + \widetilde{Y}_1^2 \mu_3^2 \mu_4^2) \nn \\
 &  + (Y_1^2 \widetilde{Y}_1^2 + Y_3^2 \widetilde{Y}_3^2) (Y_2^2 \mu_1^2 
\mu_3^2 + \widetilde{Y}_2^2 \mu_2^2 \mu_4^2) \nn \\
 &  + (Y_1^2 \widetilde{Y}_1^2 + Y_2^2 \widetilde{Y}_2^2) (Y_3^2 \mu_1^2 
\mu_4^2 + 
\widetilde{Y}_3^2 \mu_2^2 \mu_3^2).\label{Xidef}
\end{align}

\section{Uplifting gauged STU supergravity to eleven dimensions} \label{sec:uplift}

The uplift ansatz for $A_{mnp}$, given in \eqref{A} in the previous section, 
along with the previous results of ref.~\cite{cvlupostutry}, allows
us to complete the  
uplift of gauged STU supergravity to eleven dimensions.  The uplift of
the four-dimensional metric $ds_4^2$ was obtained in ref.~\cite{cvlupostutry}, and
is given by
\bea
d\hat s_{11}^2 &=& \Xi^{\fft13}\, ds_4^2 + \Xi^{\fft13}\, d\hat s_7^2
\nn\\
&=& \Xi^{\fft13}\, ds_4^2 + g^{-2}\, \Xi^{-\fft23}\, 
\Big[ \sum_\alpha Z_\alpha\, (d\mu_\alpha^2 + \mu_\alpha^2\,
D\phi_\alpha^2) + 2b_2\, b_3\, (\mu_1^2\, \mu_2^2\, D\phi_1\, D\phi_2 
- \mu_3^2\, \mu_4^2\, D\phi_3\, D\phi_4)\nn\\
&&+ 2b_1\, b_3\, (\mu_1^2\, \mu_3^2 \,D\phi_1\, D\phi_3 
- \mu_2^2\, \mu_4^2\, D\phi_2\, D\phi_4) 
+ 2b_1\, b_2\, (\mu_1^2\, \mu_4^2\, D\phi_1\, D\phi_4 
- \mu_2^2\, \mu_3^2\, D\phi_2\, D\phi_3)\nn\\
&&+\ft12 b_1^2\, \Big( (\mu_1\, d\mu_1 + \mu_2\, d\mu_2)^2 + 
                     (\mu_3\, d\mu_3 + \mu_4\, d\mu_4)^2\Big)\nn\\
&&+\ft12 b_2^2\, \Big( (\mu_1\, d\mu_1 + \mu_3\, d\mu_3)^2 + 
                     (\mu_2\, d\mu_2 + \mu_4\, d\mu_4)^2\Big)\nn\\
&&+\ft12 b_3^2\, \Big( (\mu_1\, d\mu_1 + \mu_4\, d\mu_4)^2 + 
                     (\mu_2\, d\mu_2 + \mu_3\, d\mu_3)^2\Big)\Big]\,,
\label{d11metricans}
\eea
where $ds_4^2$ is the four-dimensional metric,
the $Z_\alpha$ are defined in (\ref{Widef}) and (\ref{Zdef}), and 
$\Xi$ is defined in (\ref{Xidef}). The 1-forms $D\phi_\alpha$,
which are defined by 
\be
D\phi_\alpha \equiv d\phi_\alpha - g\, A_\1^\alpha\,,
\ee
manifest the expected $U(1)^4$ gauge invariance $A_\1^\alpha(x) \longrightarrow
A_\1^\alpha(x) + d\Lambda^\alpha(x)$ that originates via the subset 
$\phi_\alpha\longrightarrow \phi_\alpha +
g\, \Lambda^\alpha(x)$ of the eleven-dimensional coordinate transformations.


The 4-form field strength is given in equation (41) of 
ref.~\cite{cvlupostutry}, which we reproduce here for convenience
\bea
 \hat{F}_{(4)} &= &-2g U \epsilon_\4  + \hat{F}'_{(4)} + 
\hat{F}''_{(4)} + \hat G_\4\,,
  \label{F}
\eea
where $\epsilon_\4$ is the volume form in the four-dimensional
spacetime metric $ds_4^2$.
The first term in (\ref{F}) corresponds to the Freund-Rubin term, with
\be
U = Y_1^2 \, (\mu_1^2+\mu_2^2) + \widetilde Y_1^2\, (\mu_3^2 + \mu_4^2) 
+ Y_2^2 \, (\mu_1^2+\mu_3^2) + \widetilde Y_2^2\, (\mu_2^2 + \mu_4^2)
+ Y_3^2 \, (\mu_1^2+\mu_4^2) + \widetilde Y_3^2\, (\mu_2^2 + \mu_3^2)\,.
\label{Udef}
\ee 
In terms of the components $F_{ABCD}$, 
this corresponds to $F_{\alpha \beta \gamma \delta}$.  

The next term in the expression above is $\hat F_\4'= d\hat A_\3'$, where
$\hat A_\3'$ is obtained by making the replacements $d\phi_\alpha \rightarrow
D\phi_\alpha$ in the expression for the internal projection of
the 3-form potential \eqref{A}.  Thus we have~\footnote{Note that we have introduced the hatted indices $\hat\alpha$ such that
 $(\hat1 ,\hat 2,\hat3,\hat 4)=(5,6,7,8)$.}
\be \label{eq:Ahatdef}
\hat A_\3' = \ft12 A_{\alpha\hat\beta\hat\gamma}\, d\mu_\alpha\wedge 
(d\phi_\beta-g\, A_\1^\beta)\wedge (d\phi_\gamma- g \, A_\1^\gamma)\,,
\ee
where the components $A_{\alpha\hat\beta\hat\gamma}$ are given in
\eqref{A}.

  The term $\hat{F}''_\4$, which  is given in equation (43) of 
ref.~\cite{cvlupostutry}, is
\be
\hat F_\4'' = - \fft1{2 g^2}\, |W|^{-2} \, \sum_\alpha d\mu_\alpha^2\wedge 
(d\phi_\alpha - g\, A_\1^\alpha)\wedge R_\alpha\,, \label{fppans}
\ee
where
\bea
R_1 &=& 
\tY_1^2\, \tY_2^2\, \tY_3^2\, [P_0\, {*F_\2^1} + 2 b_1\, b_2\, b_3\, F_\2^1] 
+\tY_1^2\, 
[P_1\, b_2\, b_3\, {*F_\2^2} + b_1\,(P_0 + 2 b_2^2\, b_3^2)\, F_\2^2]
\label{ridef}\\
&&
+\tY_2^2 \,
[P_2\, b_1\, b_3\, {*F_\2^3} + b_2\,(P_0 + 2 b_1^2\, b_3^2)\, F_\2^3]
+\tY_3^2 \,
[P_3\, b_1\, b_2\, {*F_\2^4} + b_3\,(P_0 + 2 b_1^2\, b_2^2)\, F_\2^4]
\,,\nn\\
R_2 &=&
\tY_1^2\, Y_2^2\, Y_3^2\, [P_0\, {*F_\2^2} + 2 b_1\, b_2\, b_3\, F_\2^2] 
+\tY_1^2\, 
[P_1\, b_2\, b_3\, {*F_\2^1} + b_1\,(P_0 + 2 b_2^2\, b_3^2)\, F_\2^1]
\nn\\
&&
-Y_2^2 \,
[P_2\, b_1\, b_3\, {*F_\2^4} + b_2\,(P_0 + 2 b_1^2\, b_3^2)\, F_\2^4]
-Y_3^2 \,
[P_3\, b_1\, b_2\, {*F_\2^3} + b_3\,(P_0 + 2 b_1^2\, b_2^2)\, F_\2^3]
\,,\nn\\
R_3 &=&
Y_1^2\, \tY_2^2\, Y_3^2\, [P_0\, {*F_\2^3} + 2 b_1\, b_2\, b_3\, F_\2^3] 
-Y_1^2\, 
[P_1\, b_2\, b_3\, {*F_\2^4} + b_1\,(P_0 + 2 b_2^2\, b_3^2)\, F_\2^4]
\nn\\
&&
+\tY_2^2 \,
[P_2\, b_1\, b_3\, {*F_\2^1} + b_2\,(P_0 + 2 b_1^2\, b_3^2)\, F_\2^1]
-Y_3^2 \,
[P_3\, b_1\, b_2\, {*F_\2^2} + b_3\,(P_0 + 2 b_1^2\, b_2^2)\, F_\2^2]
\,,\nn\\
R_4 &=&
Y_1^2\, Y_2^2\, \tY_3^2\, [P_0\, {*F_\2^4} + 2 b_1\, b_2\, b_3\, F_\2^4] 
-Y_1^2 \,
[P_1\, b_2\, b_3\, {*F_\2^3} + b_1\,(P_0 + 2 b_2^2\, b_3^2)\, F_\2^3]
\nn\\
&&
-Y_2^2 \,
[P_2\, b_1\, b_3\, {*F_\2^2} + b_2\,(P_0 + 2 b_1^2\, b_3^2)\, F_\2^2]
+\tY_3^2\,
[P_3\, b_1\, b_2\, {*F_\2^1} + b_3\,(P_0 + 2 b_1^2\, b_2^2)\, F_\2^1]
\,,\nn
\eea
and $W$ and $P_\alpha$ are defined in (\ref{WPdef}).  

   Finally, the remaining term in \eqref{F} is given by
\bea
\hat G_\4 &=& \fft1{2g}\, (2 Y_1^{-1}\, {*d Y_1} - \chi_1\, Y_1^{4}\, 
        {* d\chi_1})\wedge d(\mu_1^2+\mu_2^2) \nn\\
&&
 +
  \fft1{2g}\, (2 Y_2^{-1}\, {*d Y_2} - \chi_2\, Y_2^{4}\,
        {* d\chi_2})\wedge d(\mu_1^2+\mu_3^2) \nn\\
&&
+ \fft1{2g}\, (2 Y_3^{-1}\, {*d Y_3} - \chi_3\, Y_3^{4}\,
        {* d\chi_3})\wedge d(\mu_1^2+\mu_4^2)\,.\label{Gdef}
\eea
Note that, as observed in ref.~\cite{cvlupostutry}, the equations of motion
for the $U(1)$ gauge fields in the STU theory are simply given by
$d(|W|^{-2}\, R_\alpha)=0$.

\subsection{Consistency of the other uplift formulae with the general uplift ans\"atze}


In this section, we revisit the uplift ans\"atze for some of the other components of the 4-form field strength. In particular, in light of the recent developments \cite{4formans}, we show that the ansatz for, in particular, the Freund-Rubin term takes a simple form.

In ref.~\cite{4formans}, an uplift ansatz is given for the Freund-Rubin term and it is, moreover, conjectured that the term can be expressed in terms of the sum of the potential and its derivative according to the conjectured formula \eqref{frconjans}. In order to express the Freund-Rubin term in this form,
 we introduce the following paramaterisation of the scalars:
\be
\eta_i = \lambda_i\, \cos\sigma_i\,,\qquad \zeta_i = \lambda_i\, \sin\sigma_i\,.
\ee
The scalars $\eta_i$ are the coefficients of the self-dual tensors in the
parameterisation of $\phi_{ijkl}$ in the unitary gauge, and $\zeta_i$
are the coefficients of the anti-self dual tensors.

    It can now be seen that the function $U$ given in (\ref{Udef}) can be
written as
\be
U= -\fft1{8 g^2}\, ( V + \sum_i \fft{\del V}{\del \eta_i}\, \xi_i)\,,
\ee
where 
\be
\xi_1= \mu_1^2 + \mu_2^2 - \mu_3^2 -\mu_4^2\,,\qquad
\xi_2= \mu_1^2 + \mu_3^2 - \mu_2^2 -\mu_4^2\,,\qquad
\xi_3= \mu_1^2 + \mu_4^2 - \mu_2^2 -\mu_3^2\,.
\ee

Note also that, in terms of tensors $\xi_i,$ $\hat G_\4$ given in (\ref{Gdef}) can be written as
\be
\hat G_\4 = \fft1{4g}\, \sum_i (\cos\sigma_i\, {*d\lambda_i} -
   \sin\sigma_i\, \sinh\lambda_i\, \cosh\lambda_i\, {*d\sigma_i})\wedge d\xi_i
\,.
\ee
The form of this is consistent with the ansatz for the $F_{\alpha \beta \gamma a}$ component of the field strength proposed in ref.~\cite{4formans}, equation \eqref{eq:F31ansatzs7}.
Furthermore, it is now straightforward to see that the Bianchi 
identity $d\hat F_\4=0$ 
is satisfied provided that the scalar fields satisfy their equations of
motion.

The $\hat{F}'_{(4)}$ term is given by the Kaluza-Klein covariantisation of $F_{mnpq}$, $F_{\mu mnp}$ and, since $\hat{F}'_{(4)} = d \hat A'_{(3)}$ with $\hat A'_{(3)}$ given by equation \eqref{eq:Ahatdef}, the second term on the right-hand side of the expression for $F_{\alpha \beta a b},$ \eqref{eq:F22s7}. 

Moreover, the $\hat{F}''_{(4)}$ is given by the Kaluza-Klein covariantisation of the first term on the right-hand side of the expression for $F_{\alpha \beta a b},$ \eqref{eq:F22s7}. 

Thus, we have established the direct, if somewhat intricate, relationship between the uplift ans\"atze of the gauged STU supergravity with the general ans\"atze for the full ${\cal N}=8$ supergravity, given in ref.~\cite{4formans}.

\section{Consistent truncations of the STU embedding}

   There are two inequivalent 
consistent truncations of STU supergravity that are 
sometimes useful in their own right, and have the merit of being considerably
simpler than the full STU theory.  In each case, the bosonic
sector of the truncated supergravity comprises gravity coupled to 
two $U(1)$ gauge fields, a dilatonic
scalar and an axionic scalar.  The two truncations, which we shall refer
to as the $2+2$ truncation and the $3+1$ truncation, are implemented
by setting
\bea
2+2:\qquad&&\lambda_1=\lambda\,\qquad \sigma_1=\sigma\,,\qquad
 \lambda_2=\lambda_3=\sigma_2=\sigma_3=0\,,\nn\\
\qquad &&
A^{1}_\mu=A^{2}_\mu=   A_\mu\,,\qquad
A^{3}_\mu=A^{4}_\mu=  \wtd A_\mu\,,\label{22trunc}\\
&&\nn\\
3+1:\qquad&& \lambda_1=\lambda_2=\lambda_3=\lambda\,,\qquad 
\sigma_1=\sigma_2=\sigma_3=\sigma\,,\nn\\
\qquad && A^{1}_\mu= \wtd A_\mu\,,\qquad
      A^{2}_\mu= A^{3}_\mu=A^{4}_\mu=  A_\mu\,.
\label{31trunc}
\eea
Note that we are not rescaling the gauge potentials in these truncations,
so $A_\mu$ and $\wtd A_\mu$ in the $2+2$ truncation and $A_\mu$ in the
$3+1$ truncation will have non-canonically normalised kinetic terms.  Also,
in the
$3+1$ truncation the scalar fields $\lambda$ and $\sigma$ will have 
non-canonically normalised kinetic terms.  We have chosen not to 
rescale the truncated fields in order to avoid
the occurrence of many $\sqrt2$ or $\sqrt3$ factors.  It will be 
convenient 
to parameterise the scalar fields in the standard $SL(2,\R)$ form,
for which their kinetic terms are proportional to $-\del\varphi^2 -
e^{2\varphi}\, \del\chi^2$, 
by introducing $\varphi$ and $\chi$ that are related to $\lambda$ and
$\sigma$ by
\be
e^\varphi= \cosh\lambda + \sinh\lambda\, \cos\sigma\,,\qquad 
\chi\, e^\varphi= \sinh\lambda\, \sin\sigma\,,
\ee
in terms of which we have
\be
Y^2= e^\varphi\,,\qquad \wtd Y^2= e^{-\varphi} +\chi^2\, e^{\varphi}
\,,\qquad
   b= \chi\, e^\varphi\,.
\ee

\subsection{$2+2$ truncation}

  Here, with the fields truncated as in (\ref{22trunc}), we choose an
adapted parameterisation for the
the $\mu_\alpha$ coordinates in which we take
\be
\mu_1= c\, \cos\ft12\theta\,,\qquad \mu_2= c\, \sin\ft12\theta\,,\qquad
\mu_3= s\, \cos\ft12\tilde\theta\,,\qquad \mu_4= s\, \sin\ft12\tilde\theta\,,
\ee
where we have also defined
\be
c=\cos\xi\,,\qquad s=\sin\xi\,.\label{scdef}
\ee
The four azimuthal angles $\phi_\alpha$ will be parameterised by defining
\be
\phi_1=\ft12(\psi+\phi)\,,\qquad \phi_2=\ft12(\psi-\phi)\,,\qquad
\phi_3=\ft12(\wtd\psi+\td\phi)\,,\qquad \phi_4=\ft12(\wtd\psi-\td\phi)\,.
\ee
The unit 7-sphere metric then takes the form
\be
d\Omega_7^2 = \sum_{\alpha=1}^4 (d\mu_\alpha^2 + \mu_\alpha^2\, d\phi_\alpha^2) 
=d\xi^2 + \cos^2\xi\, d\Omega_3^2 + \sin^2\xi\, d\wtd\Omega_3^2\,,
\ee
where
\be
d\Omega_3^2 = \ft14\Big[d\theta^2 + \sin^2\theta\, d\phi^2 +
    (d\psi+\cos\theta\, d\phi)^2\Big]\,,\quad
d\wtd\Omega_3^2 = \ft14\Big[d\td\theta^2 + \sin^2\td\theta\, d\td\phi^2 +
    (d\wtd\psi+\cos\td\theta\, d\td\phi)^2\Big]\,
\ee
are the metrics on two unit 3-spheres.

   The metric reduction ansatz (\ref{d11metricans}) in the $2+2$ truncation
thus reduces to
\bea
d\hat s_{11}^2 &=& \Xi^{\ft13}\, ds_4^2 + \fft{\Xi^{\ft13}}{g^2}\,\Big\{
d\xi^2 +
  \fft{\cos^2\xi}{4 Z_3}\, \Big[d\theta^2 + \sin^2\theta\, d\phi^2 +
 (d\psi + \cos\theta\, d\phi- 2g A_\1)^2\Big] \nn\\
&& \qquad\qquad\qquad\quad +\fft{\sin^2\xi}{4 Z_1}\, \Big[d\tilde\theta^2 + 
  \sin^2\tilde\theta\, d\tilde\phi^2 +
 (d\wtd\psi + \cos\tilde\theta\, d\tilde\phi- 2 g \wtd A_\1)^2\Big]\Big\}\,,
\eea
where, from (\ref{Widef}) and (\ref{Zdef}), we now have
\bea
Z_1&=&Z_2= \wtd Y^2\, \sin^2\xi   + \cos^2\xi\,,\quad
Z_3=Z_4= Y^2\,  \cos^2\xi + \sin^2\xi\,,\quad
\Xi= Z_1\, Z_3\,.\label{2+2met}
\eea

  The 4-form reduction ansatz is now given by 
\bea
\hat F_\4 &=& -2 g\, U\, \epsilon_\4 -
   \fft{ s c}{g}\,( {*d\varphi} - \chi\, e^{2\varphi}\,
   {*d\chi})\wedge d\xi + d\hat A_\3' + \hat F_\4''\,,\nn\\
\hat A_\3' &=& \fft1{g^3}\, \chi\, e^{\varphi} \,
\Big[\fft{c^4}{Z_3}\, \Omega(A) - \fft{s^4}{Z_1}\, \wtd\Omega(\wtd A)\Big]\,,
\\
\hat F_\4'' &=& \fft{c}{2 g^2 Y^2}\,
   \Big[s d\xi\wedge (d\psi +\cos\theta\, d\phi -2 g A_\1) +
   \ft12 c\, \sin\theta\, d\theta\wedge d\phi\Big]\wedge ({*F}_\2 + 
   \chi\, Y^2\, F_\2) \nn\\
&& \!\!\!\!\! - \fft{s}{2 g^2 \wtd Y^2} \, \Big[
 c\, d\xi \wedge (d\wtd\psi + \cos\tilde\theta\, d\tilde\phi - 2g
    \wtd A_\1) -\ft12 s\, \sin\tilde\theta\, d\tilde\theta\wedge 
   d\tilde\phi\Big]\wedge ({*\wtd F}_\2 - \chi\, Y^2\, \wtd F_\2)\,,\nn
\eea
where
\bea
U&=& c^2\, Y^2 + s^2\, \wtd Y^2 + 2\,,\nn\\
\Omega(A) &=& \ft18 \sin\theta\, 
(d\psi +\cos\theta\, d\phi - 2 g A_\1)\wedge d\theta\wedge d\phi\,,\nn\\
\wtd\Omega(\wtd A) &=& \ft18 \sin\tilde\theta\, 
(d\wtd\psi +\cos\tilde\theta\, d\tilde\phi - 2 g \wtd A_\1)
\wedge d\tilde\theta\wedge d\tilde\phi\,.
\eea

   The bosonic sector of the 
STU supergravity Lagrangian given in section 3 reduces under the
present $2+2$ truncation to
\bea
{\cal L}&=& R\, {*\oneone} - \ft12 {*d\varphi}\wedge d\varphi -
  \ft12 e^{2\varphi}\, {*d\chi}\wedge d\chi - V\, {*\oneone} \nn\\
&& -  Y^{-2}\, {*F}_\2\wedge F_\2 - \wtd Y^{-2}\, {*\wtd F}_\2
\wedge \wtd F_\2 \nn\\
&&- \chi\, F_\2\wedge F_\2 +  \chi Y^2\, \wtd Y^{-2}
\, \wtd F_\2\wedge \wtd F_2\,,
\eea
where
\be
V= -4 g^2\, (Y^2+\wtd Y^2 + 4)\,.
\ee

  It can be verified that the uplift of the $2+2$ truncation that we obtained 
in this subsection agrees with the abelian truncation of the uplift of
the ${\cal N}=4$ gauged $SO(4)$ supergravity that was obtained in 
ref.~\cite{cvluposo4}.

\subsection{$3+1$ truncation}

  For the $3+1$ truncation of the fields of STU supergravity, given by 
equation (\ref{31trunc}), an appropriate adapted parameterisation for the 
$\mu_\alpha$ coordinates is given by taking
\be
\mu_1=\cos\xi\,,\qquad \mu_a= \nu_a\, \sin\xi\,,\quad a=2,3,4\,,\qquad
\sum_a \nu_a^2=1\,.
\ee
The unit $S^7$ metric then takes the form
\be
d\Omega_7^2 = d\xi^2 + \cos^2\xi\, d\phi_1^2 + \sin^2\xi\, d\Omega_5^2\,,
\ee
where
\be
d\Omega_5^2 = \sum_{a=2}^4 (d\nu_a^2 +\nu_a^2\, d\phi_a^2) 
   = d\Sigma_2^2 + (d\psi + B)^2
\ee
is the metric on the unit 5-sphere, and $d\Sigma_2^2$ is the standard
``unit'' Fubini-Study metric\footnote{The Fubini-Study metric
is given in terms of the complex coordinates $z_a=\nu_a\, e^{\im \phi_a}$
on $S^5$ by $d\Sigma_2^2= \sum_a d\bar z_a dz_a -|\sum_a \bar z_a dz_a|^2$.
This ``unit'' metric is Einstein with $R_{ij}=6 g_{ij}$.} 
on $\CP^2$, with $J=\ft12 dB$ being the
K\"ahler form on $\CP^2$.  In terms of $\nu_a$ and $\phi_a$ we have
\be
(d\psi + B) =  \sum_a \nu_a^2\, d\phi_a\,.
\ee

   It will be convenient to define the two functions
\be
\beta= Y^2\, (Y^2\, c^2 + \wtd Y^2\, s^2)\,,
\qquad \gamma=Y^4\, c^2 + s^2\,,
\label{betagammadef}
\ee
where, as before, we have defined $c=\cos\xi$ and $s=\sin\xi$. 
 From (\ref{Widef}), (\ref{Zdef}) and (\ref{Xidef}) we find
\be
Z_1= \wtd Y^4 \, s^2 + c^2\,,\qquad Z_a=\beta - b^2\, s^2\, \nu_a^2\,,
\qquad \Xi= \beta^2\, Y^{-2}\,,
\ee
where, as before, $b=\chi\, e^{\varphi}= \chi\, Y^2$.   The 
eleven-dimensional metric (\ref{d11metricans}) now becomes, in the
$3+1$ truncation we are considering here,
\bea
d\hat s_{11}^2 &=& \Xi^{\fft13}\, ds_4^2 + g^{-2}\, \Xi^{-\fft23}\,
\Big[\fft{\beta^2}{Y^4}\, d\xi^2 + \gamma\, s^2\, 
\Big((d\psi + B - g A_\1) + \fft{b^2\, c^2}{\gamma}\, 
 (d\phi_1- g \wtd A_\1)\Big)^2 \nn\\
&&\qquad\qquad\qquad\qquad
+\beta\, s^2\, d\Sigma_2^2 +
  \fft{\beta^2\, c^2}{\gamma\, Y^4}\, (d\phi_1-g \wtd A_\1)^2\Big]\,.
\eea

   Substituting the $3+1$ truncation into the uplift formula for the
4-form in STU supergravity given in section 5, we now find
\be
\hat F_\4 = -2 g U\, \epsilon_\4 + \hat G_\4 + d\hat A_\3' +\hat F_\4''\,,
\ee
with 
\bea
U &=& 2 (Y^2\, c^2 + \wtd Y^2\, s^2) + Y^2\,,\nn\\
\hat G_\4 &=& -\fft{2sc}{g}\, ({*d\varphi} - \chi\, e^{2\varphi}\, 
{*d\chi})\wedge d\xi\,,\nn\\
\hat A_\3' &=& \fft{sc\, \chi}{g^3}\, d\xi\wedge (d\phi_1-g\wtd A_\1) \wedge
  (d\psi + B -g A_\1) - \fft{s^2 c^2}{\beta g^3}\, \chi\, e^{2\varphi}\,
(d\phi_1-g \wtd A_\1)\wedge J \nn\\
&& +\fft{s^4}{\beta g^3}\, \chi\, e^{2\varphi}\,
(d\psi+ B-g A_\1)\wedge J\,,\nn\\
\hat F_\4''&=& \fft{sc}{g^2\, |W|^2}\, d\xi\wedge \wtd R\wedge 
  (d\phi_1-g\wtd A_\1) -\fft{sc}{g^2\, |W|^2}\, d\xi\wedge R\wedge
  (d\psi + B- g A_\1)\nn\\
&&- \fft{s^2}{g^2\, |W|^2}\, R\wedge J\,,
\eea
where, from (\ref{WPdef}), 
\be
|W|^2= (1+4b^2)(1+b^2)^2
\ee
and from (\ref{ridef}), 
\be
\wtd R= R_1\,, \qquad R=R_2=R_3=R_4\,,
\ee
with
\bea
\wtd R &=& \wtd Y^6\, [(1+3b^2)\, {*\wtd F_\2} + 2 b^3\, \wtd F_\2] +
   3 b\, (1+b^2)\, \wtd Y^2\, [b \, {*F_\2} + (1+2b^2)\, F_\2]\,,\nn\\
R &=& Y^2\, (1+b^2)^2\, [{*F}_\2 - 2 b\, F_\2] + 
   b\, (1+b^2)\, \wtd Y^2\, [b\, {*\wtd F_\2} + (1+2b^2)\, \wtd F_\2]\,.
\eea

  The bosonic sector of the STU supergravity Lagrangian given in section 3 
reduces, under the $3+1$ truncation, to
\be
{\cal L}_4 = R\, {*\oneone} - \ft32 
\left ({*d\vp} \wedge d\vp + e^{2\vp}\, {*d\chi} \wedge
d\chi \right) -V\, {*\oneone} + {\cal L}_{KinA} + {\cal L}_{CS} \,,
\ee
with
\bea
V&=&-12 g^2 (Y^2+ \wtd Y^2), \nn\\
{\cal L}_{KinA} &=& -\fft 1{2(1+4\chi^2\,e^{2\vp})} \, \big[6 \chi^2\,e^{\vp} 
\, {*F_\2} \wedge\wtd F_\2+ 
 e^{-3\vp}(1+3\chi^2 e^{2\vp})(1+\chi^2e^{2\vp})\,  {*\wtd F_\2} \wedge 
\wtd F_\2 \nn\\
&&\qquad\qquad\qquad\qquad
  +3 e^{\vp} \, {*F_\2}\, \wedge F_\2\big], \nn\\
{\cal L}_{CS} &=& -\fft{\chi}{(1+4\chi^2\,e^{2\vp})}\, \big[-3e^{2\vp} 
F_\2\, \wedge \,F_\2 \,+ 3(1+2\chi^2\,e^{2\vp}) \, \wtd F_\2\wedge F_\2
 \nn\\
&& \qquad\qquad\qquad\qquad 
   +\chi^2 (1+\chi^2\,e^{2\vp})\, \wtd F_\2\, \wedge \wtd F_\2\big].
\eea

\section{Conclusions}

The existence of a consistent
reduction of eleven-dimensional supergravity on $S^7$, to give 
four-dimensional ${\cal N}=8$
$SO(8)$ gauged supergravity, was first established in \cite{dewitnics7}.  In
that paper, the complete expression for the eleven-dimensional metric
ansatz was obtained, and partial expressions also for the four-form
field strength.  The complete expressions for the uplifted four-form
field strength were obtained recently \cite{dWN13,gengeo,godgodnic,4formans}.
Although the results are now complete, they are not necessarily 
convenient to use in practice, if, as is commonly the case, one
is interested in uplifting four-dimensional configurations that involve
only a subset of the full set of ${\cal N}=8$ supergravity fields.  In the
context, for example, of four-dimensional black hole solutions, it is 
almost always the case that only an abelian subsector of the 
$SO(8)$ gauge fields is turned on.  Thus for many practical purposes,
it suffices to know the uplift formulae for the truncation of the 
$SO(8)$ gauged supergravity to its abelian $U(1)^4$ subsector.  The
consistent truncation of the maximal gauged supergravity to this abelian 
subsector corresponds to the gauged ${\cal N}=2$ STU supergravity theory, 
whose bosonic sector comprises the metric, the four abelian $U(1)$ gauge
fields, three dilatonic scalars and three axionic scalars.  This theory
is, for example, sufficient in order to describe the general class of
rotating asymptotically AdS black holes, which can carry four electric and
four magnetic charges.

In this paper, we have carried out the abelian truncation of the results 
for the uplifting of the gauged ${\cal N}=8$ theory, thereby obtaining 
fully explicit expressions for the
lifting of the bosonic sector of the four-dimensional gauged STU supergravity
to eleven dimensions.  We formulated the truncation in a notation that
is adapted to the $U(1)^4$ isometries in the internal directions on 
the seven-sphere.  In special cases where the gauge fields are either 
purely electric or purely magnetic, one can make a further consistent
truncation (to a non-supersymmetric theory) in which the three axionic
scalars vanish.  This leads to enormous simplifications in the formulae,
and they then reduce to ones that are given in ref.~\cite{tenauthor}.

We also considered two distinct supersymmetric truncations of the
STU supergravity, where, in each case, two gauge fields, a dilatonic
scalar and an axionic scalar survive.  The first, which we referred to
as the $2+2$ truncation, is achieved by setting the four original gauge
fields to be pairwise equal.  The second, which we
call the $3+1$ truncation, is achieved instead by setting three of the
original four gauge fields equal.  The geometric structure of the 
internal seven-sphere becomes particularly simple in these two
truncations.  In the $2+2$ truncation, the seven-sphere is described
as a foliation by $S^3\times S^3$ factors, with the two surviving gauge
fields being associated with the two $U(1)$ isometries acting on the
Hopf fibres in the two $S^3$ factors, viewed as $U(1)$ bundles over $S^2$.
In the $3+1$ truncation the seven-sphere is instead described as a foliation by
$S^5\times S^1$ factors, with the two surviving gauge fields being associated
with the $U(1)$ isometry acting on the Hopf fibres of $S^5$ viewed as
a $U(1)$ bundle over $\CP^2$, and the $U(1)$ isometry of the $S^1$ factor.  
Consequently, the uplift formulae are much simpler for these truncations.

We hope that the new uplift ans\"atze presented in this paper for what are
particularly interesting truncations of maximal $SO(8)$ gauged supergravity
will be of use in future applications.

\section*{Acknowledgements}

We would like to thank Olaf Kr\"uger and Hermann 
Nicolai for discussions. H.G.\ and M.G.\ thank the Mitchell Institute 
for Fundamental Physics and Astronomy, Texas A$\&$M University
for hospitality during the course of this work.  H.G., M.G.\ and
C.N.P.\ also thank the AEI, Golm, and the Mitchell Family Foundation for 
hospitality at
the Great Brampton House workshop.  
H.G.\ and M.G.\ are supported by King's College, 
Cambridge. H.G.\ acknowledges funding
from the European Research Council under the European Community's
Seventh Framework Programme (FP7/2007-2013) / ERC grant agreement no.
[247252].  C.N.P.\ is
supported in part by DOE grant DE-FG02-13ER42020.

\appendix
 
\section{$U(1)^4$-invariant tensors in adapted coordinates} \label{app:coords}

We shall use an adapted coordinate system  $y^m=(\mu_\alpha,\phi_\rho)$ 
for $S^7$, with the direction cosines $\mu_\alpha$ satisfying
\begin{equation}
 \sum_{\alpha=1}^{4} \mu_\alpha^2= 1.
\end{equation}
These coordinates are related to the $\R^8$ coordinates as follows
\be
x^1+\im x^2 = \mu_1\, e^{\im\phi_1}\,,\quad
x^3+\im x^4 = \mu_2\, e^{\im\phi_2}\,,\quad
x^5+\im x^6 = \mu_3\, e^{\im\phi_3}\,,\quad
x^7+\im x^8 = \mu_4\, e^{\im\phi_4}\,.\label{xmuphi}
\ee
The expressions for the various $U(1)^4$-invariant tensors become much more elegant in
terms of these coordinates.  First of all, we
have the background round $S^7$ metric
\be
d\Omega_7^2 = \go_{mn} dy^m dy^n = \sum_\alpha(d\mu_\alpha^2 + \mu_\alpha^2\, d\phi_\alpha^2)\,.
\ee
Note that the inverse background metric in these coordinates is
\begin{equation}
\po_{7}^2 = \go^{mn} \partial_{y^m} \partial_{y^n} = 
\sum_{\alpha < \beta} (\mu_\alpha \partial_{\mu_\beta} - \mu_\beta 
\partial_{\mu_\alpha})^2 + \sum_\alpha \mu_\alpha^{-2}\, 
\partial^2_{\phi_\alpha}\,.
\end{equation}

The vectors $\xi^{(i)}_m$, viewed as 1-forms, are then given by
\begin{eqnarray}
\xi^{(1)}_m dy^m &=& \ft32 d(\mu_1^2 + \mu_2^2)= 
3(\mu_1\, d\mu_1 + \mu_2\, d\mu_2) 
= -\ft32 d(\mu_3^2 + \mu_4^2)=
-3(\mu_3\, d\mu_3 + \mu_4\, d\mu_4)
\,,\nn\\
\xi^{(2)}_m dy^m &=& \ft32 d(\mu_1^2 + \mu_3^2)= 
3(\mu_1\, d\mu_1 + \mu_3\, d\mu_3)
= -\ft32 d(\mu_2^2 + \mu_4^2)=
-3(\mu_2\, d\mu_2 + \mu_4\, d\mu_4)
\,,\nn\\
\xi^{(3)}_m dy^m &=& \ft32 d(\mu_1^2 + \mu_4^2)= 
3(\mu_1\, d\mu_1 + \mu_4\, d\mu_4)
= -\ft32 d(\mu_2^2 + \mu_3^2)=
-3(\mu_2\, d\mu_2 + \mu_3\, d\mu_3)\,. \nn \\
\end{eqnarray}

  The 2-index symmetric tensors $\xi^{(i)}_{mn}$ are given by
\begin{align}
 \xi^{(1)}_{mn}\, dy^m dy^n = &-\ft14 (\mu_1^2+\mu_2^2)(d\mu_1^2+ d\mu_2^2 + \mu_1^2 d\phi_1^2 +\mu_2^2 d\phi_2^2) \nn \\
 &\hspace{25mm} + \ft14 (\mu_3^2+\mu_4^2)(d\mu_3^2+ d\mu_4^2 + \mu_3^2 d\phi_3^2 +\mu_4^2 d\phi_4^2)\,,\nn\\
 \xi^{(2)}_{mn}\, dy^m dy^n = &-\ft14 (\mu_1^2+\mu_3^2)(d\mu_1^2+ d\mu_3^2 + \mu_1^2 d\phi_1^2 +\mu_3^2 d\phi_3^2) \nn \\
 &\hspace{25mm} + \ft14 (\mu_2^2+\mu_4^2)(d\mu_2^2+ d\mu_4^2 + \mu_2^2 d\phi_2^2 +\mu_4^2 d\phi_4^2)\,,\nn\\
 \xi^{(3)}_{mn}\, dy^m dy^n = &-\ft14 (\mu_1^2+\mu_4^2)(d\mu_1^2+ d\mu_4^2 + \mu_1^2 d\phi_1^2 +\mu_4^2 d\phi_4^2) \nn \\
 &\hspace{25mm} + \ft14 (\mu_2^2+\mu_3^2)(d\mu_2^2+ d\mu_3^2 + \mu_2^2 d\phi_2^2 +\mu_3^2 d\phi_3^2)\,.
\end{align}

  The vectors $F^{(\alpha)}_m$ are given by
\bea
F^{(1)}_m dy^m &=& \mu_1^2\, d\phi_1 + \mu_2^2\, d\phi_2 + \mu_3^2\, d\phi_3
   +\mu_4^2\, d\phi_4\,,\nn\\
F^{(2)}_m dy^m &=& \mu_1^2\, d\phi_1 + \mu_2^2\, d\phi_2 - \mu_3^2\, d\phi_3
   -\mu_4^2\, d\phi_4\,,\nn\\
F^{(3)}_m dy^m &=& \mu_1^2\, d\phi_1 - \mu_2^2\, d\phi_2 + \mu_3^2\, d\phi_3
   -\mu_4^2\, d\phi_4\,,\nn\\
F^{(4)}_m dy^m &=& \mu_1^2\, d\phi_1 - \mu_2^2\, d\phi_2 - \mu_3^2\, d\phi_3
   +\mu_4^2\, d\phi_4\,.
\eea
The 2-forms $F^{(\alpha)}_\2  = d F^{(\alpha)}_\1$ are given by
\bea
\ft12 F^{(1)}_{mn} dy^m\wedge dy^n &=&  2\mu_1 d\mu_1\wedge d\phi_1 +
  2\mu_2 d\mu_2\wedge d\phi_2 + 2\mu_3 d\mu_3\wedge d\phi_3 +
   2\mu_4 d\mu_4\wedge d\phi_4\,,\nn\\
\ft12 F^{(2)}_{mn} dy^m\wedge dy^n &=&  2\mu_1 d\mu_1\wedge d\phi_1 +
  2\mu_2 d\mu_2\wedge d\phi_2 - 2\mu_3 d\mu_3\wedge d\phi_3 -
   2\mu_4 d\mu_4\wedge d\phi_4\,,\nn\\
\ft12 F^{(3)}_{mn} dy^m\wedge dy^n &=&  2\mu_1 d\mu_1\wedge d\phi_1 -
  2\mu_2 d\mu_2\wedge d\phi_2 + 2\mu_3 d\mu_3\wedge d\phi_3 -
   2\mu_4 d\mu_4\wedge d\phi_4\,,\nn\\
\ft12 F^{(4)}_{mn} dy^m\wedge dy^n &=&  2\mu_1 d\mu_1\wedge d\phi_1 -
  2\mu_2 d\mu_2\wedge d\phi_2 - 2\mu_3 d\mu_3\wedge d\phi_3 +
   2\mu_4 d\mu_4\wedge d\phi_4\,.\nn\\
\eea

The 3-forms $S^{(i)}$ are given by
\bea
S^{(1)} &=& -\ft12 \mu_1\mu_2\, (\mu_1 d\mu_2-\mu_2 d\mu_1)
   \wedge d\phi_1\wedge d\phi_2 + 
   \ft12 \mu_3\mu_4\, (\mu_3 d\mu_4-\mu_4 d\mu_3)
   \wedge d\phi_3\wedge d\phi_4\,,\nn\\
S^{(2)} &=& -\ft12 \mu_1\mu_3\, (\mu_1 d\mu_3-\mu_3 d\mu_1)
   \wedge d\phi_1\wedge d\phi_3 +
   \ft12 \mu_2\mu_4\, (\mu_2 d\mu_4-\mu_4 d\mu_2)
   \wedge d\phi_2\wedge d\phi_4\,,\nn\\
S^{(3)} &=& -\ft12 \mu_1\mu_4\, (\mu_1 d\mu_4-\mu_4 d\mu_1)
   \wedge d\phi_1\wedge d\phi_4 +
   \ft12 \mu_2\mu_3\, (\mu_2 d\mu_3-\mu_3 d\mu_2)
   \wedge d\phi_2\wedge d\phi_3\,.\nn\\
\eea

\bibliographystyle{utphys}
\bibliography{stu}

\providecommand{\href}[2]{#2}\begingroup\raggedright\begin{thebibliography}{10}

\bibitem{dewitnicsg}
B.~de~Wit and H.~Nicolai, ``{$N=8$ supergravity},''
\href{http://dx.doi.org/10.1016/0550-3213(82)90120-1}{{\em Nucl.Phys.} {\bf
  B208} (1982)  323}.

\bibitem{dufpop}
M.~J. Duff and C.~N. Pope, ``Kaluza-{K}lein supergravity and the
  seven-sphere,'' pp.~183--228.
\newblock World Scientific, 1983.

\bibitem{biendeni}
B.~Biran, F.~Englert, B.~de~Wit, and H.~Nicolai, ``{Gauged $N=8$ supergravity
  and its breaking from spontaneous compactification},''
  \href{http://dx.doi.org/10.1016/0370-2693(83)91400-4}{{\em Phys. Lett.} {\bf
  B124} (1983)  45}.
[Erratum: Phys. Lett.B128,461(1983)].

\bibitem{frerub}
P.~G.~O. Freund and M.~A. Rubin, ``Dynamics of dimensional reduction,''
\href{http://dx.doi.org/10.1016/0370-2693(80)90590-0}{{\em Phys.Lett.} {\bf
  B97} (1980)  233--235}.

\bibitem{gibbonshat}
G.~W. Gibbons, {public communication}.

\bibitem{dunipowa}
M.~J. Duff, B.~E.~W. Nilsson, C.~N. Pope, and N.~P. Warner, ``{On the
  consistency of the {Kaluza-Klein} ansatz},''
\href{http://dx.doi.org/10.1016/0370-2693(84)91558-2}{{\em Phys. Lett.} {\bf
  B149} (1984)  90}.

\bibitem{pauli}
W.~Pauli, {\em {Wissenschaftlicher Briefwechsel, Vol.~IV, Part II}}.
\newblock Springer-Verlag, 1999.

\bibitem{Straumann:2000zc}
N.~Straumann, ``{On Pauli's invention of non-Abelian Kaluza-Klein theory in
  1953},'' in {\em {Recent developments in theoretical and experimental general
  relativity, gravitation and relativistic field theories. Proceedings, 9th
  Marcel Grossmann Meeting, MG'9, Rome, Italy, July 2-8, 2000. Pts. A-C}},
  pp.~1063--1066.
\newblock 2000.
\newblock
\href{http://arxiv.org/abs/gr-qc/0012054}{{\tt arXiv:gr-qc/0012054}}.
\newblock

\bibitem{ORaifeartaigh:1998pk}
L.~O'Raifeartaigh and N.~Straumann, ``{Early history of gauge theories and
  Kaluza-Klein theories, with a glance at recent developments},''
\href{http://arxiv.org/abs/hep-ph/9810524}{{\tt arXiv:hep-ph/9810524}}.

\bibitem{cvgilupo}
M.~Cveti{\v c}, G.~W. Gibbons, H.~L{\"u}, and C.~N. Pope, ``{Consistent group
  and coset reductions of the bosonic string},''
  \href{http://dx.doi.org/10.1088/0264-9381/20/23/013}{{\em Class. Quant.
  Grav.} {\bf 20} (2003)  5161--5194},
\href{http://arxiv.org/abs/hep-th/0306043}{{\tt arXiv:hep-th/0306043}}.

\bibitem{dewitnics7}
B.~de~Wit and H.~Nicolai, ``{The consistency of the {$S^7$} truncation in
  $D=11$ supergravity},''
\href{http://dx.doi.org/10.1016/0550-3213(87)90253-7}{{\em Nucl.Phys.} {\bf
  B281} (1987)  211}.

\bibitem{nicpil}
H.~Nicolai and K.~Pilch, ``{Consistent truncation of $d = 11$ supergravity on
  $AdS_4 \times S^7$},'' \href{http://dx.doi.org/10.1007/JHEP03(2012)099}{{\em
  JHEP} {\bf 1203} (2012)  099},
\href{http://arxiv.org/abs/1112.6131}{{\tt arXiv:1112.6131}}.

\bibitem{dWN13}
B.~de~Wit and H.~Nicolai, ``{Deformations of gauged $SO(8)$ supergravity and
  supergravity in eleven dimensions},''
\href{http://arxiv.org/abs/1302.6219}{{\tt arXiv:1302.6219}}.

\bibitem{gengeo}
H.~Godazgar, M.~Godazgar, and H.~Nicolai, ``{Generalised geometry from the
  ground up},'' \href{http://dx.doi.org/10.1007/JHEP02(2014)075}{{\em JHEP}
  {\bf 1402} (2014)  075},
\href{http://arxiv.org/abs/1307.8295}{{\tt arXiv:1307.8295}}.

\bibitem{godgodnic}
H.~Godazgar, M.~Godazgar, and H.~Nicolai, ``{Nonlinear Kaluza-Klein theory for
  dual fields},'' \href{http://dx.doi.org/10.1103/PhysRevD.88.125002}{{\em
  Phys. Rev.} {\bf D88} (2013) no.~12, 125002},
\href{http://arxiv.org/abs/1309.0266}{{\tt arXiv:1309.0266}}.

\bibitem{4formans}
H.~Godazgar, M.~Godazgar, O.~Kr{\"u}ger, and H.~Nicolai, ``{Consistent 4-form
  fluxes for maximal supergravity},''
  \href{http://dx.doi.org/10.1007/JHEP10(2015)169}{{\em JHEP} {\bf 10} (2015)
  169},
\href{http://arxiv.org/abs/1507.07684}{{\tt arXiv:1507.07684}}.

\bibitem{cvluposo4}
M.~Cveti{\v c}, H.~L{\"u}, and C.~Pope, ``{Four-dimensional $N=4$, $SO(4)$
  gauged supergravity from $D = 11$},''
  \href{http://dx.doi.org/10.1016/S0550-3213(99)00828-7}{{\em Nucl.Phys.} {\bf
  B574} (2000)  761--781}, \href{http://arxiv.org/abs/hep-th/9910252}{{\tt
  arXiv:hep-th/9910252}}.

\bibitem{tenauthor}
M.~Cveti{\v c}, M.~J. Duff, P.~Hoxha, J.~T. Liu, H.~L{\"u}, J.~X. Lu,
  R.~Martinez-Acosta, C.~N. Pope, H.~Sati, and T.~A. Tran, ``{Embedding AdS
  black holes in ten dimensions and eleven dimensions},''
  \href{http://dx.doi.org/10.1016/S0550-3213(99)00419-8}{{\em Nucl. Phys.} {\bf
  B558} (1999)  96--126},
\href{http://arxiv.org/abs/hep-th/9903214}{{\tt arXiv:hep-th/9903214}}.

\bibitem{cvlupostutry}
M.~Cveti{\v c}, H.~L{\"u}, and C.~Pope, ``{Geometry of the embedding of
  supergravity scalar manifolds in $D = 11$ and $D = 10$},''
  \href{http://dx.doi.org/10.1016/S0550-3213(00)00215-7}{{\em Nucl.Phys.} {\bf
  B584} (2000)  149--170},
\href{http://arxiv.org/abs/hep-th/0002099}{{\tt arXiv:hep-th/0002099}}.

\bibitem{testing}
H.~Godazgar, M.~Godazgar, and H.~Nicolai, ``{Testing the non-linear flux ansatz
  for maximal supergravity},''
  \href{http://dx.doi.org/10.1103/PhysRevD.87.085038}{{\em Phys.Rev.} {\bf D87}
  (2013)  085038},
\href{http://arxiv.org/abs/1303.1013}{{\tt arXiv:1303.1013}}.

\bibitem{duffliu}
M.~J. Duff and J.~T. Liu, ``{Anti-de Sitter black holes in gauged $N = 8$
  supergravity},'' \href{http://dx.doi.org/10.1016/S0550-3213(99)00299-0}{{\em
  Nucl. Phys.} {\bf B554} (1999)  237--253},
\href{http://arxiv.org/abs/hep-th/9901149}{{\tt arXiv:hep-th/9901149}}.

\bibitem{dWNW85}
B.~de~Wit, H.~Nicolai, and N.~P. Warner, ``The embedding of gauged ${N}=8$
  supergravity into $d = 11$ supergravity,''
\href{http://dx.doi.org/10.1016/0550-3213(85)90128-2}{{\em Nucl.Phys.} {\bf
  B255} (1985)  29}.

\bibitem{Varela:2015ywx}
O.~Varela, ``{The complete $D=11$ embedding of SO(8) supergravity},''
\href{http://arxiv.org/abs/1512.04943}{{\tt arXiv:1512.04943 [hep-th]}}.

\bibitem{Kruger:2016agp}
O.~Kr{\"u}ger, ``{Non-linear uplift Ansätze for the internal metric and the
  four-form field-strength of maximal supergravity},''
  \href{http://dx.doi.org/10.1007/JHEP05(2016)145}{{\em JHEP} {\bf 05} (2016)
  145},
\href{http://arxiv.org/abs/1602.03327}{{\tt arXiv:1602.03327 [hep-th]}}.

\bibitem{Lu:2014fpa}
H.~L{\"u}, Y.~Pang, and C.~N. Pope, ``{An $\omega$ deformation of gauged STU
  supergravity},'' \href{http://dx.doi.org/10.1007/JHEP04(2014)175}{{\em JHEP}
  {\bf 04} (2014)  175},
\href{http://arxiv.org/abs/1402.1994}{{\tt arXiv:1402.1994}}.

\end{thebibliography}\endgroup

\end{document}